\documentclass[10pt]{iopart}

\usepackage{graphicx}
\usepackage{amsthm}
\usepackage{amssymb}
\usepackage{eepic}
\usepackage{amscd}
\usepackage{longtable}
\usepackage{array}
\usepackage{graphicx}
\usepackage{slashbox}
\usepackage{multirow} 
\usepackage{color} 
\usepackage{bigdelim,multirow}

\newcommand{\dst}{\displaystyle}

\newcommand{\bcsp}{\!\!\!\!}

\newcommand{\bfa}{{\mathbf a}}

\newcommand{\cX}{{\mathcal X}}
\newcommand{\A}{{\mathcal A}}
\newcommand{\T}{{\mathcal T}}

\newcommand{\N}{{\mathbb N}}

\newcommand{\C}{{\mathbb C}}
\newcommand{\one}{\mbox{1 \hspace{-3.2mm} {\bf \rm l}} }

\newcommand{\eps}{{\epsilon}}
\newcommand{\del}{{\delta}}
\newcommand{\ot}{{\otimes}}
\newcommand{\vp}{{\varphi}}

\newcommand{\frs}{{\mathfrak s}}
\newcommand{\frt}{{\mathfrak t}}

\newcommand{\cross}{{\ -\!\!\!-\!\!\!-{\ }_{\ }
 \!\!\! \!\!\! \!\!\! \!\!\! \!\!\! \!
 {\ }_{{\ }_\lambda}\!\Big|\quad}}

\begin{document}

\title[{Generalized matrix Ansatz in the multispecies ASEP}]{Generalized matrix Ansatz in the multispecies exclusion process -- partially asymmetric case}

\author{Chikashi Arita$^{1}$,
Arvind Ayyer$^{2}$, \\
Kirone Mallick$^{1}$,
Sylvain Prolhac$^{3,4}$}

\address{
 $^{1}$ Institut de Physique Th\'eorique CEA, F-91191 Gif-sur-Yvette, France\\
 $^{2}$ University of California Davis, One Shields Avenue, Davis 95616  USA\\
 $^{3}$ Zentrum Mathematik, Technische Universit\"at M\"unchen, Germany\\
 $^{4}$ Department of Physics of Complex Systems, Weizmann Institute of Science, Israel}

\ead{chikashi.arita@cea.fr, ayyer@math.ucdavis.edu,
kirone.mallick@cea.fr, prolhac@ma.tum.de}

\begin{abstract}
We investigate one of the simplest multispecies generalization of the
asymmetric simple  exclusion process on a ring. This process has a
rich combinatorial spectral structure and  a matrix product form for
the stationary state.  In the totally asymmetric case operators that
conjugate the dynamics of systems with different numbers of species
were obtained by  the authors  and reported recently \cite{AAMP1}.
The existence of such nontrivial  operators was reformulated as a
representation problem for a  specific quadratic algebra (generalized
matrix Ansatz).  In the present work, we construct the  family of
representations    explicitly  for  the  partially
asymmetric  case.  This solution cannot  be
obtained   by a simple deformation  of the totally  asymmetric case.
\end{abstract}

\maketitle

\section{Introduction}

The fundamental quest of statistical mechanics is to derive macroscopic behavior
from  microscopic laws. In this respect,  
the  asymmetric simple exclusion process (ASEP) has been playing 
a central role for the last two decades \cite{DerrCairns,Varadhan1}. The ASEP
 is one of the few models for which the  hydrodynamic limit has been  proved mathematically, 
using elaborate  large deviation techniques \cite{Varadhan,Spohn}.  In
nonequilibrium statistical mechanics, the  exclusion process, 
 being   one of the simplest non-trivial  interacting particle process that one can imagine,
  has reached  the status of a paradigm  \cite{Li99,Spitzer}. The ASEP is  a lattice-gas where
 each particle is a  (possibly  biased)
 random walker hopping from a site  to one of the neighboring  locations 
  only if the target site is empty. 
  This exclusion constraint mimics a hard-core interaction
 amongst particles. If the hopping rates are not isotropic, a non-vanishing flow
 of particles is transported through the system which is therefore permanently  driven out
of equilibrium. 
 Despite its simplicity, the ASEP displays a 
 deep   mathematical structure that  has enabled many researchers to 
perform precise analytical studies. These results  can be used as benchmarks in the ongoing
construction of a general theory of nonequilibrium systems 
 \cite{BE,DerridaRep,DerrMaths,DerrCairns,ogkmrev,KMCairns,PaulK,SZ,Schutz2}.

 The standard  exclusion process involves  particles and holes hopping
 on a one-dimensional lattice, although many variants have been studied 
 \cite{BE,Schutz2}.
 A  straightforward   generalization of the ASEP is to consider 
 the  case of multiple species  of particles with hierarchical dynamical rules
 \cite{AB,AKSS,AR,PEM,Wehefritz}. 
 In fact, such a model  with $N$ different species of particles automatically  appears
 when one couples $N$ standard  exclusion processes \cite{Liggett} in a natural 
 way. The $N$-ASEP provides a fundamental example of a multicomponent
 non-equilibrium process; it has highly non-trivial steady states,
which are not Gibbs states in general 
and depend on boundary conditions
 \cite{Arita,BE,PEM,Speer}.  

For the
 totally  asymmetric  case  ($N$-TASEP),
  the stationary state was 
 constructed  combinatorially  by Ferrari and Martin \cite{FM}. This construction
 was restated  as a  matrix product  Ansatz in \cite{EFM}
 and  was generalized in \cite{PEM} 
 to the partially asymmetric case ($N$-PASEP).

In a recent study \cite{AAMP1} of the $N$-TASEP, 
 the matrix product  Ansatz
 was generalized in order to construct
 a conjugation operator that embeds
 the $(N-1)$-TASEP in  the  $N$-TASEP.
By considering the whole family of
 $N$-TASEP processes,
 with varying $N$, a network of mappings can be  constructed (corresponding to an underlying
 partially ordered set --{\it poset}--  structure). It was shown that the 
 Ferrari and Martin  construction was a special case of a more general algorithm, corresponding
 to a  generalized matrix Ansatz that allows one to lift information from a system containing less
 species of particles to a system containing more  species, by recursively splitting 
 identical classes  of particles into different species. Moreover, the  information that is obtained 
 is  not restricted to the steady states but also  affects subsets of the spectrum.
 However, the  results of  \cite{AAMP1} were only valid for the  $N$-TASEP: the general 
 Ferrari and Martin  algorithm  cannot directly be applied to the $N$-ASEP model.
 Besides, the 
 representations of
  the  generalized matrix Ansatz that were obtained for the  $N$-TASEP could not
 be deformed to the $N$-ASEP in general. 
 The purpose of the present work is to fill this gap: we shall explain how to construct 
 all  the conjugation operators by providing an explicit representation for the quadratic algebras
 involved in the  $N$-ASEP poset structure. We shall also explain how the  $N$-ASEP  relates to the  Perk-Schultz model and investigate the relationship between
  the conjugation operators  and the Perk-Schultz 
  transfer matrix as well.

 The outline of the present  work is as follows.
 In Section~\ref{sec:Review}, 
 we define the model, discuss its basic features  and
  briefly review  the spectral inclusion  properties
  and the generalized matrix Ansatz.
  The Ansatz starts with a local
  relation which gives a quadratic algebra.
The representation
 for this algebra in the totally  asymmetric   case  \cite{AAMP1},
 however, cannot be extended directly
 to the  partially asymmetric   case  (the $N$-PASEP).
We will present a family of representations
  for the PASEP  in Section \ref{sec:solPASEP}.
In Section \ref{sec:Perk}, we explain the relevance of our results
to the  Perk-Schultz model.
We present the conclusion 
of the present work in Section \ref{sec:conclusion}.

\section{Known results about the $N$-ASEP model}
\label{sec:Review}

\subsection{Definition of the  model}

Consider a ring ${\mathbb Z}_L$ with $L$ sites,
 where a variable
  (local state) $k_i\in\{1,\dots,N+1\}$ 
 is assigned to each site $i\in \mathbb Z_L$.
Nearest neighbor pairs of local states 
$JK$
are interchanged $J K \to K J$
with the transition rate
\begin{eqnarray}\label{eq:rate}
\Theta(J-K) = \left\{
\begin{array}{cc}
p& {\rm if }\ J < K,\\
q& {\rm if }\ J > K
\end{array}\right.
\end{eqnarray}
with $\Theta(0)=0$.
Without loss of generality,
we set $0\le q\le p=1$.
We say that the site $i$ is occupied by
 a $J$th class particle if $k_i=J\le N$.
We regard the site $i$ as being empty if $k_i=N+1$.
When  $q=0$, the model is totally asymmetric and is called the $N$-TASEP;
for $0<q<1$, the model is  partially  asymmetric  and is called the $N$-PASEP.
The special case $q=1$, corresponds to the symmetric simple exclusion process (SSEP).

The dynamics of the $N$-ASEP can be encoded in a master equation. 
Let $\{|1\rangle, \ldots,|N+1\rangle\}$ be the basis of the 
single-site space $\C^{N+1}$ and
$|k_1\dots k_L\rangle $ be the tensor product
$|k_1\rangle \ot\cdots \ot |k_L\rangle
\in (\C^{N+1})^{\ot L}$.
In terms of the probability vector
\begin{eqnarray}
 |P(t)\rangle = \sum_{1\le k_i \le N+1}
 P(k_1\cdots k_L; t)|k_1 \cdots k_L\rangle,
\end{eqnarray}
the master equation that governs the system 
 is expressed as 
\begin{eqnarray}\label{eq:master2}
\frac{d}{dt}|P(t)\rangle = M^{(N)}|P(t)\rangle.
\end{eqnarray}
The linear operator $ M^{(N)}$ has the form
\begin{eqnarray}\label{eq:tot-Markov}
 M^{(N)}&= \sum_{i \in {\mathbb Z}_L}
\left(M^{(N)}_{\rm Loc}\right)_{i, i+1}
\label{eq:k:hdef}
\end{eqnarray}
where  the local operators 
$\left(M^{(N)}_{\rm Loc}\right)_{i, i+1}$ act 
on the $i$th and the $(i+1)$th components
of the tensor product and are given by 
\begin{eqnarray}
  M^{(N)}_{\rm Loc} = 
  \sum_{J,K=1}^{N+1}
   \left(-\Theta(J-K)|JK\rangle\langle JK|
      +\Theta(J-K)|KJ\rangle\langle JK|\right). 
\label{eq:MLoc}
\end{eqnarray}
The Markov matrix has zero as an eigenvalue
and the corresponding eigenvector is called the
stationary state.  For a given number  $m_i\in \mathbb N$ of particles of type
$i$ (with $ 1 \le i \le N+1$), the stationary state is unique. 
All the other eigenvalues have 
strictly negative real parts, which characterize
the relaxation to the stationary states.
The  Markov matrix defines an integrable model that can be solved
 by means of the nested algebraic Bethe Ansatz  \cite{AB,AKSS}.
Besides, in \cite{AKSS},
the spectral structure of the   $N$-ASEP   Markov matrix  was  investigated; 
one simple result is that  the spectrum of   the    Markov matrix  
of $N$ species of particles 
contains  the spectra  of systems with  $N'<N$ species.

\subsection{Matrix Ansatz for the stationary state}

 Although the  $N$-ASEP is integrable by Bethe Ansatz, the  explicit  calculation of eigenvectors
is not easy even for the stationary states.
 An alternative technique for studying  the stationary state is the matrix 
 product Ansatz
 (which we call simply matrix Ansatz), first used in \cite{DEHP}.
 This trick has grown into very powerful 
 method to analyze one-dimensional systems out of equilibrium \cite{BE, DerrMaths}. 
 The matrix  Ansatz for the  stationary state of the   $N$-ASEP was constructed in
 \cite{EFM, PEM}. 
The basic idea is to write  the stationary weight for a configuration
$k_1\cdots k_L$ as a trace of a matrix product
\begin{equation}
\label{eq:matrix-stationary}
   P(k_1\cdots k_L) = \frac{1}{Z} \Tr (X_{k_1}\cdots X_{k_L}),
\end{equation}
where the operator $X_k$ corresponds to
 the $k$-th species and $Z$ is a normalization
 factor.
 The   operators  $ X_k  (k=1,\dots,N+1)$ must  obey  specific  relations
 in order for 
 the expression  (\ref{eq:matrix-stationary})
to represent the stationary state. 
 We emphasize that  the representation space of
 these operators  $X_k$ is not  the physical
 space but an abstract vector space which is usually infinite dimensional.
 It was shown  in  \cite{EFM, PEM}  that  the  $X_k$'s can be chosen as sums of 
 tensor products 
 of $\del_q,\eps_q,A_q$ and $\one$,
 which satisfy the relations
\begin{eqnarray}
\del_q \eps_q -q\eps_q \del_q = (1-q)\one,\quad
\del_q A_q = qA\del_q,\quad
A_q \eps_q =q\eps_q A_q \, . 
\label{q-algebra}
\end{eqnarray}
This algebra is related to 
the quantum harmonic oscillator and the quantum group
 \cite{Arita, Macfarlane,Sa,Schutz1}.
An explicit representation of this  quadratic  algebra is given, for example, by 
the  following infinite dimensional matrices \cite{DEHP}:
\begin{eqnarray}\label{eq:fund-mat}
\fl\quad
\eqalign{
 \del_q =
\left(\begin{array}{cccc}
  0   &  c_1   &      &   \\
       &   0  &   \!\! c_2 &    \\[-2mm]
       &      &   \!\! 0  & \!\! \ddots  \\[-2mm]
       &      &      &  \!\! \ddots
\end{array}\!\!\!\right),\ 
 \eps_q=
\left(\begin{array}{cccc}
   0   & &      &   \\
   c_1 &   0  &  &   \\
       & c_2  & \!\! 0  & \\[-2mm]
      &      & \!\!\ddots &   \!\!\ddots
\end{array}\!\!\!\right) ,\ 
A_q=\left(\begin{array}{cccc}
1  &  &      &     \\
&  q & &          \\
  &  &  q^2 &     \\[-2mm]
    &      &  &   \!\!\ddots
\end{array}\!\!\!\right) ,
}
\end{eqnarray}
and $\one = A_1$
with $c_i=\sqrt{1-q^i}$.

\subsection{Poset structure and spectral inclusion}\label{sec:poset}

\begin{figure}[h]
  \begin{center}
 \includegraphics[height=7cm]{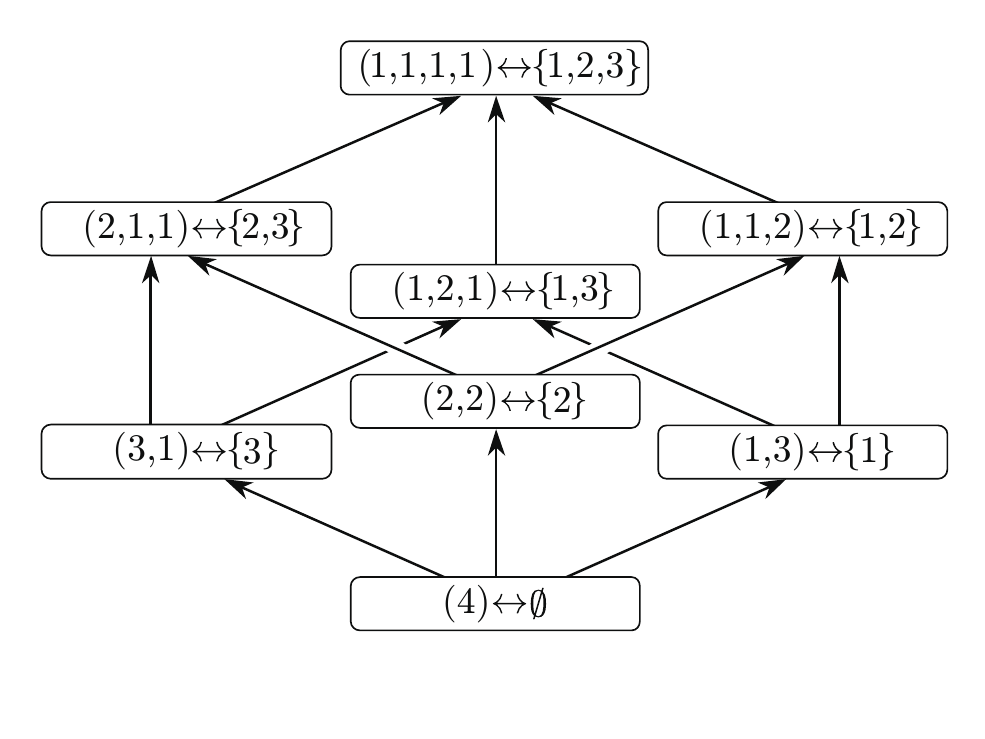} 
 \caption{The Hasse diagram for $L=4$. Each basic sector is labelled in two ways.}
  \label{fig:Hasse}
  \end{center}
 \end{figure}

Since the number of each class of particles is conserved, 
  the total Markov matrix
  $M^{(N)}$ (\ref{eq:tot-Markov})
  splits into blocks as
\begin{equation}
\label{eq:=otMm}
\fl\quad
M^{(N)}=  \bigoplus_{m}M_m, \,\,\,  \hbox{ with } 
   \,\,\, \ M_m \in {\rm End} V_m,\,\,\,  \hbox{ where}   \,\,\, 
V_m= \bigoplus_{\# \{i|k_i=j \} = m_j} \mathbb C |k_1\cdots k_L \rangle \,  .
\end{equation}
We labeled each diagonal block $M_m$
and the corresponding vector space (sector) 
$V_m$ by $m=(m_1,\dots,m_{N+1})$.
We will call the label itself a sector,
and in particular, a {\it basic sector} corresponds to the case  
$m_i > 0 $ for all $i$. A useful  alternative
labeling of the  basic sectors \cite{AKSS} is obtained as  follows:  
let $s_j$ be the total number of particles of classes $k \le j$, 
\begin{eqnarray}\label{eq:one-to-one}
s_j = m_1+ m_2 + \cdots + m_j.
\end{eqnarray}
One has $m_j = s_j-s_{j-1} > 0$  
($s_0=0$),
and thus each basic sector
 can be labeled by the set
  $\frs = \{s_1,  \dots, s_N\} \subset 
    \{1,2,\dots, L-1\} =\Omega $
 with $0<s_1 < s_2 < \cdots < s_N<L$.
The set $\frs$ is an
 element of ${\mathcal S}$, the power set (the set of all
 subsets) of $ \Omega.$
In the following, we shall use both labels equivalently:
 for instance, the invariant vector spaces (respectively  the  Markov
 matrices acting on them)  will be denoted either by $V_m$ or $V_\frs$
 (respectively   $M_m$ or  $M_\frs$).
The set ${\mathcal S}$ is equipped
with a natural poset (partially ordered set)
structure  with respect to the inclusion $\subseteq$,
 which is encoded in the Hasse diagram.
In our case it is 
 simply the $L-1$ dimensional hypercube
where each vertex of the  hypercube  corresponds  to a
 sector,
 and each edge corresponds to  an arrow
$\frt\rightarrow \frs$  meaning that $\frt \subset \frs$ and 
$\#\mathfrak{s} = \# \frt + 1$.
(See figure \ref{fig:Hasse}.)

\hfill\break

\hfill\break

The  spectral properties of the Markov matrix
   on the Hasse diagram were investigated  in \cite{AKSS}. Given two sectors
 $\frs=\{s_1<\cdots<s_N\}$
and $\frt=\frs\setminus\{s_{n_1},\dots,s_{n_u}\}$ connected 
 by a finite sequence  of arrows in the Hasse diagram, one
 could define an   identification operator from $\frs$ to $\frt$ as follows 
\begin{eqnarray}\label{eq:defphi}
\fl\quad
 \varphi_{\frt\frs} : \ 
| k_1 \cdots k_L\rangle \in V_\mathfrak{s}
 \ \mapsto\ 
| k'_1 \cdots k'_L\rangle \in V_\mathfrak{t}\quad {\rm with}
  \quad x' = x-\#\{i| n_i <x\} \, . 
\end{eqnarray}
Using  $\varphi_{\frt\frs}$,
 a {\it conjugation relation} could be proved: 
 $ \varphi_{\frt\frs} M_\frs = M_\frt  \varphi_{\frt\frs}.$
This relation implies that by applying   $\varphi_{\frt\frs}$, 
  an eigenvector $|E\rangle_\frs$ of the sector $\frs$
 with an eigenvalue $E$ is either  projected to 
 an eigenvector $\varphi_{\frt \frs}|E\rangle_\frs$
 in sector $\frt$ or is  killed out.
The surjectivity of $\varphi_{\frt\frs}$ leads to 
the spectral inclusion
 $ {\rm Spec}  (M_\frs)   \supset  {\rm Spec}  (M_\frt) \, .$

\subsection{Generalized matrix Ansatz and the hat-algebra}

It should be noted that the action of
$\varphi_{\frt\frs}$ loses information by projecting a larger sector $\frs$ into a smaller one $\frt$.
What is desirable is a {\it conjugation operator} ({\it conjugation matrix})
$\psi_{\frs\frt}:V_{\frt}\to V_{\frs}$ satisfying
the opposite conjugation relation 
\begin{equation}\label{eq:main-relation}
   M_\frs \psi_{\frs\frt}  = \psi_{\frs\frt} M_\frt  \, , 
\end{equation}
which can be expressed
 by the following commutative diagram:
\begin{eqnarray} \label{DiagComm2}
\begin{CD}
  V_\frs  @> \dst M_\frs >>  V_\frs    \\
  @A \dst \psi_{\frs\frt} AA  @AA \dst \psi_{\frs\frt} A  \\
  V_\frt  @> \dst M_\frt >>  V_\frt
\end{CD}
\end{eqnarray}
The form for $\psi_{\frs\frt}$ is nontrivial in general,
 and its action helps us to construct eigenvectors
(including stationary states) from lower sectors.

In \cite{AAMP1} a method to
construct the conjugation matrix
  $\psi_{\frs\frt}$ was introduced
  by generalizing the matrix Ansatz
  for the stationary state.
The basic idea is to write
each element of $\psi_{\frs\frt}$ 
in a form 
\begin{eqnarray}
\label{eq:def-psi-ele}
  \langle j_1\cdots j_L |
     \psi_{\frs\frt} | k_1\cdots k_L \rangle
 =\Tr (a_{j_1k_1}\cdots a_{j_Lk_L}) \,  
\end{eqnarray}
with matrices $a_{JK}$
$(1\le J\le N+1, 1\le K\le N).$

 Consider for a nearest-neighbor pair 
  of sectors
   $\frs=\{s_1<\cdots<s_N\}$
   of  $N$-species     and
   $\frt = \frs\setminus\{s_n\}$
 of  $(N-1)$-species.
  The sector  $\frt$  is  
  obtained from the sector  $\frs$ 
  by merging
  the particles of $n$th and
   $(n+1)$st classes  into the  same class. 
It will be convenient to write  these operators as elements of an operator-valued matrix $\bfa$ such that 
\begin{eqnarray}
a_{JK} = \langle J| \bfa|K\rangle \, .
\end{eqnarray}
  Following \cite{AAMP1}, one can prove
  that the matrix
  $\psi_{\frs,\frs\setminus\{s_n\}}$
  defined by  equation~(\ref{eq:def-psi-ele})  is a conjugation operator 
 if there  exists  another operator-valued matrix $\widehat \bfa^{(N,n)}$
 of size  $N+1$ by $N$  that satisfies
  the following
 key relation  ({\it hat relation}):
\begin{eqnarray}\label{eq:hat-relation}
 M^{(N)}_{\rm Loc} ({\bfa \ot \bfa })
  -   ({\bfa  \ot \bfa })  M^{(N-1)}_{\rm Loc}   =
  {\bfa  \ot\widehat \bfa  - \widehat \bfa  \ot \bfa } \,.
\end{eqnarray}
The elements of this equation 
give   a set of relations,
which we call the  {\it hat algebra}.  
We emphasize that the hat relation~(\ref{eq:hat-relation})  connects 
 the local Markov matrix of the $N$-ASEP and that of the $(N-1)$-ASEP,
and is 
of a  different nature from 
 the relations  used  in \cite{HSP,Rajewsky} for 
 the usual matrix Ansatz for  the  stationary state.
 
    Therefore, the problem of constructing   a conjugation operator
 has been reduced to finding realizations  of the  hat algebra.

 \subsection{The $N$-TASEP case}

In the TASEP case, 
  a family of solutions
  $(\bfa,\widehat \bfa)
   = (\bfa^{(N,n)},\widehat \bfa^{(N,n)})$
    ($1\le n\le N$)
 to the hat relation (\ref{eq:hat-relation}) was successfully
  constructed  in \cite{AAMP1},
where  $n$ is the class being split.
The operator-valued matrices
${\bfa^{(N,n)}}$ and ${ \widehat \bfa^{(N,n)} }$ are related {by multiplication with a diagonal matrix}:
\begin{eqnarray}
\label{eq:TASEP-hat}
{\widehat \bfa^{(N,n)}} ={\rm diag} 
(\underbrace{1,\dots,1}_{n},\underbrace{0,\dots,0}_{N+1-n}){\bfa^{(N,n)}} \, .
\end{eqnarray}
 Then the hat algebra~(\ref{eq:hat-relation}) becomes a closed  quadratic algebra
 generated by the elements of  ${\bfa^{(N,n)}}$. An explicit representation
 of the hat algebra is constructed 
 by $(N-1)$-fold
  tensor products of the fundamental quadratic algebra generated
 by the infinite dimensional matrices 
$\del,\eps$ and $A$ that satisfy the relations \cite{DEHP}
\begin{eqnarray}
\label{eq:TASEP-fund-rel}
 \del \, \eps = 1 \,, \,\,\,  \del \,  A  = 0 \,, \,\,\,  A \, \eps = 0 \, .
\end{eqnarray}
 We note that 
they are obtained   by taking  $q=0$ in
 (\ref{eq:fund-mat}).

The expressions of the operators $ a^{(N,n)}_{JK}$  are given in the following table with
  {\rm $\del\ot\one^{\ot(-1)}\ot\eps=\one$}
and {\rm $\del\ot\one^{\ot x}\ot\eps=0$ for $x\le -2$}.

\begin{eqnarray}
\fl\quad
\begin{array}{|c|c|c|c|}
\hline 
\ & \ & \ & \  \\[-8pt]
{}_{\displaystyle J}\diagdown K
    & 1\quad \cdots\quad n-1 & n& n+1\quad \cdots\quad N
\\[2pt] \hline
 \begin{array}{cc}1\\[-3pt] \vdots\\[-3pt] n-1 \end{array}
 \bcsp&\bcsp
  \begin{array}{cc} A^{\ot (J-1)}\ot\del\ot\one^{\ot(K-J-1)} \\
                  \ot\eps\ot\one^{\ot(N-K-1)} \end{array}
 \bcsp&\bcsp
  \begin{array}{cc} A^{\ot(J-1)} \ot\\ \del \ot\one^{\ot(N-J-1)} \end{array}
 \bcsp&\bcsp
  \begin{array}{cc} A^{\ot(J-1)} \ot\del\ot\one^{\ot(K-J-2)} \\
    \ot\del\ot\one^{\ot(N-K)} \end{array}
\\[2pt] \hline \ & \ & \ & \  \\[-8pt]
 n \bcsp&\bcsp 0 \bcsp&\bcsp
  \begin{array}{cc} A^{\ot (n-1)} \ot\\ \one^{\ot(N-n)} \end{array}
 \bcsp&\bcsp
   \begin{array}{cc} A^{\ot(n-1)} \ot\one^{\ot(K-n-1)} \\
    \ot\del \ot\one^{\ot(N-K)}\end{array}
\\[2pt] \hline \ & \ & \ & \  \\[-8pt]
 n+1
 \bcsp&\bcsp
  \begin{array}{cc} \one^{\ot(K-1)}\ot\eps\ot\\
    \one^{\ot(n-K-1)} \ot A^{\ot(N-n)} \end{array}
 \bcsp&\bcsp
    \begin{array}{cc} \one^{\ot (n-1)}  \ot\\ A^{\ot(N-n)} \end{array}
 \bcsp&\bcsp 0
\\[2pt] \hline
\begin{array}{cc}n+2\\[-3pt] \vdots\\[-3pt] N+1 \end{array}
 \bcsp&\bcsp
  \begin{array}{cc} \one^{\ot (K-1)} \ot\eps\ot\one^{\ot(J-K-3)} \\
                  \ot\eps\ot A^{\ot(N-J+1)} \end{array}
 \bcsp&\bcsp
  \begin{array}{cc} \one^{\ot(J-3)}\ot\\ \eps\ot A^{\ot(N-J+1)}  \end{array}
 \bcsp&\bcsp
  \begin{array}{cc} \one^{\ot(K-2)} \ot\del\ot\one^{\ot(J-K-2)} \\
    \ot\eps\ot A^{\ot(N-J+1)} \end{array}
\\[2pt] \hline
\end{array}
\label{eq:TASEP-solution}
\end{eqnarray}

\section{Generalized matrix Ansatz for the $N$-PASEP}
\label{sec:solPASEP}

  In this section, we explain how to construct 
  representations of the hat algebra
 and the corresponding conjugation operator for
 the $N$-PASEP case.
 It turns out that finding a representation  for the  hat algebra of the $N$-PASEP
 is not a simple deformation of the  $N$-TASEP case. It  will require
 a  construction  more involved and rather
 different from  that used for  the $N$-TASEP.

\subsection{The  hat relations for PASEP}

For any given value of $N$,
the TASEP solution (\ref{eq:TASEP-solution}) can be
generalized to the PASEP   
in the two special  cases $n=1$ and $n=N$.
The  case $n=1$  corresponds to
  splitting the first-class particles in two sub-classes.
The dual case $n=N$ 
 corresponds to
  splitting the holes (labelled as  $N$'s) 
  in the  $(N-1)$-PASEP  model 
   into  $N$th-class particles
   and holes (now labelled as $(N+1)$'s) 
   in the  $N$-PASEP  model.
The generalization 
 is obtained by the replacement
  $\{\del,\eps,A\}
  \to \{\del_q,\eps_q,A_q\}$ (\ref{eq:fund-mat})
in each element of $\bfa^{(N,n)}$
and a simple modification 
of (\ref{eq:TASEP-hat})
 \begin{eqnarray}
{  \widehat \bfa^{(N,1)} } &=& {\rm diag} (1,q,\dots,q) \,  {\bfa^{(N,1)} } \, , \label{cas:hat1}
 \\    
 {  \widehat \bfa^{(N,N)} }  &=& {\rm diag} (1,\dots,1,q) \,  {\bfa^{(N,N)} } \, ,
\label{cas:hatN}
  \end{eqnarray}
so that the hat relation (\ref{eq:hat-relation}) is satisfied.
The hat algebra of the TASEP case
is deformed by this modification.

For general  $n$ (with $1 < n < N$), 
equations~(\ref{cas:hat1}) and~(\ref{cas:hatN}) naturally lead us to assume  
  the following relation between 
 ${\widehat \bfa^{(N,n)} }$   and ${\bfa^{(N,n)} }$: 
 \begin{equation}\label{eq:ahat=diaga}
\widehat \bfa^{(N,n)}= d^{(N,n)}\bfa^{(N,n)}
={\rm diag} 
(\underbrace{1,\dots,1}_{n},\underbrace{q,\dots,q}_{N+1-n}) \, \bfa^{(N,n)} \, .
\end{equation}
 Inserting  this equation 
 in  the hat relation (\ref{eq:hat-relation}), one finds a  {\it closed } quadratic algebra
 for  the  operators $a_{JK}$.

 The quadratic hat algebra for the $N$-PASEP is defined by a set of relations
 between its generators, summarized  as\footnote{
  The cases A-II and C-II  vanish  
  when $q=0$. }
\begin{eqnarray}
\label{eq:explicit-algebra}
\fl\quad
\begin{array}{|l|c|c|}
\hline 
\ &   \  & \ \\[-8pt]
  &  {\rm I}: K<K'  & {\rm II}  :K\ge K'
\\[2pt] \hline \ &   \ & \  \\[-8pt]
{\rm A} : J\le n<J'
 &  \begin{array}{l}
        a_{JK'}a_{J'K} - q a_{J'K}a_{JK'} \\ = (1-q) a_{JK}a_{J'K'}
      \end{array}
 &     q a_{J'K}a_{JK'} = q a_{JK'}a_{J'K}
      \\[2pt] \hline  \ & \ & \  \\[-8pt]
 {\rm B}:
 \begin{array}{l}J<J'\le n\  {\rm or} \\ n<J<J' \end{array}
 &  a_{JK'}a_{J'K} = q a_{J'K}a_{JK'}
 &   \begin{array}{l}
         qa_{J'K}a_{JK'} - qa_{JK'}a_{J'K}  \\
          = (1-q) a_{JK}a_{J'K'} 
      \end{array}
\\[2pt] \hline  \ & \ & \  \\[-8pt]
 {\rm C}: J=J'
  & a_{JK}a_{JK'}= a_{JK'}a_{JK}
  & q a_{JK}a_{JK'}  =  q a_{JK'}a_{JK}
\\[2pt] \hline  \ & \ & \  \\[-8pt]
{\rm D}:  J>n\ge J'
 & a_{J'K}a_{JK'}= a_{JK'}a_{J'K}
 &  \begin{array}{l}
        a_{J'K}a_{JK'} - q a_{JK'}a_{J'K} \\ = (1-q) a_{JK}a_{J'K'}
      \end{array}
\\[2pt] \hline  \ & \ & \  \\[-8pt]
 {\rm E}: \begin{array}{l}J>J'>n\  {\rm or} \\ n\ge J>J'\end{array}
 & \begin{array}{l} a_{JK'}a_{J'K} - a_{J'K}a_{JK'}
   \\  = (1-q) a_{JK}a_{J'K'}  \end{array}
 & \begin{array}{l} a_{J'K}a_{JK'} \\ = q a_{JK'}a_{J'K}
    \end{array}
\\[2pt] \hline
\end{array}
\end{eqnarray}
 
Our aim is to construct an explicit representation for
this  $N$-PASEP algebra. At  first thought, one  would expect that the replacement
  $\{\del, \eps, A\}
  \to\{\del_q, \eps_q, A_q\}$ 
   in the TASEP solution
   (\ref{eq:TASEP-solution})
provides us with a representation of the hat algebra
 (\ref{eq:explicit-algebra})
 for any value of $n$.
Unfortunately,  this 
 is true only for $n=1$ and $n=N$.
For general $n$ ($1<n<N)$,  we found 
 no  perturbative representation 
 $a^{(N,n)}_{JK} = 
 a^{(N,n)}_{JK}  \Big|_{\rm TASEP}
 + q b^{(N,n)}_{JK}+ q^2 c^{(N,n)}_{JK}
 +\cdots $
 for the algebra (\ref{eq:explicit-algebra}) 
 starting from the TASEP representation (\ref{eq:TASEP-solution}).
 The { heuristic} reason behind this fact can be summarized as follows.
 If one considers the fundamental TASEP algebra ${\mathcal T}$  generated by   $\del, \eps$ and $A$ then
 the tensor products  $\del \ot \del$, $\eps \ot \eps$ and $A \ot A$ also generate the
 same algebra i.e. they satisfy the same quadratic relations.
Therefore there exists a simple
 coproduct  operation 
 from   ${\mathcal T}$ to  ${\mathcal T} \ot {\mathcal T}$ that preserves the algebraic structure  (\ref{eq:TASEP-fund-rel})
  \cite{Chari}.
 However, one can easily verify that  $\del_q \ot \del_q$, $\eps_q \ot \eps_q$ and $A_q \ot A_q$
 do not satisfy the   fundamental PASEP  relations~(\ref{q-algebra}). 
It seems that  no coproduct exists 
for   PASEP algebra that would allow us  to build natural tensor 
  representations.
This is the mathematical obstruction  that  prevents us from constructing
 the $N$-PASEP hat  algebra by deforming the known  $N$-TASEP hat  algebra.

 To summarize, the 
  solution for the $N$-TASEP 
   (\ref{eq:TASEP-solution})
   is of no help in general
 to find representations of the $N$-PASEP  algebra~(\ref{eq:explicit-algebra}).
 One needs a different strategy to build explicit representations,
 which we will explain
 beginning with   a specific example in the next subsection.

\subsection{The simplest non-trivial example}
 \label{subsec:32}

In this subsection, we work out  the simplest
 non-trivial example with $ 1 < n <N $: this is obtained 
 for  the case $(N,n)=(3,2)$.
In particular, we show that  the 
 relations~(\ref{eq:explicit-algebra}) are not contradictory by  giving an explicit  representation. Hence they  define a bona fide  algebra. 
 
A solution to (\ref{eq:explicit-algebra}) 
will be constructed by using 
the regular representation of this algebra.
We assume that
\begin{equation}
  a_{11}=a_{43}= Id,
  \quad a_{21}=a_{33}=0,\quad
\end{equation}
as in the TASEP solution.
We also assume that equation~(\ref{eq:ahat=diaga})  is valid
so that  $\widehat \bfa^{(3,2)} = (1,1,q,q) \, \bfa^{(3,2)} \, .$
Inserting these  assumptions  in the hat relation~(\ref{eq:hat-relation}),
 one obtains  28 relations, shown 
 in the leftmost column of  table \ref{tab:32alg}, that have to be satisfied
 by   the  8 unknown $a_{ij}$'s.
\begin{table}
\begin{center}
{\small
\begin{equation*}
\begin{array}{lll}
\hline \\[-10pt]
\mbox{\normalsize PASEP}&\mbox{\normalsize TASEP}&
\mbox{\normalsize Ordering} \\
\hline \\[-10pt]
 a_{13}a_{12}=a_{12}a_{13}  & a_{13}a_{12}=a_{12}a_{13}  
\label{eq:13-12} \\
 a_{12}a_{22}=q a_{22}a_{12}  & a_{12}a_{22}=0  &   a_{12}\prec a_{22}  
  \label{eq:12-22} \\
 a_{13}a_{22}=q a_{22}a_{13}  & a_{13}a_{22}=0  &  a_{13}\prec a_{22}
  \label{eq:13-22}   \\
 a_{23}a_{12}=a_{12}a_{23}+(1-q)a_{22}a_{13}  & a_{23}a_{12}=a_{12}a_{23}+a_{22}a_{13} 
  \label{eq:23-12}   \\
 a_{13}a_{23}=q a_{23}a_{13}  & a_{13}a_{23}=0  & a_{13}\prec a_{23}
 \label{eq:13-23}   \\
 a_{23}a_{22}=a_{22}a_{23}  & a_{23}a_{22}=a_{22}a_{23} 
   \label{eq:23-22} \\
 a_{12}a_{31}=q a_{31}a_{12}+(1-q)a_{32}  & a_{12}a_{31}=a_{32}  & 
 a_{12}\prec a_{31}
   \label{eq:12-31} \\
 a_{13}a_{31}=q a_{31}a_{13}  & a_{13}a_{31}=0  & 
 a_{13}\prec a_{31}
  \label{eq:13-31} \\
 a_{22}a_{31}=q a_{31}a_{22}  & a_{22}a_{31}=0  & a_{22}\prec a_{31}
   \label{eq:22-31} \\
 a_{23}a_{31}=q a_{31}a_{23}  & a_{23}a_{31}=0  & a_{23}\prec a_{31}
   \label{eq:23-31} \\
a_{12}a_{32}=a_{32}a_{12}  & a_{12}a_{32}=a_{32}a_{12} 
  \label{eq:12-32} \\
 a_{13}a_{32}=q a_{32}a_{13}  & a_{13}a_{32}=0  & a_{13}\prec a_{32}
   \label{eq:13-32} \\
 a_{22}a_{32}=a_{32}a_{22}  & a_{22}a_{32}=a_{32}a_{22}  
  \label{eq:22-32} \\
 a_{23}a_{32}=q a_{32}a_{23}  & a_{23}a_{32}=0  & a_{23}\prec a_{32}
   \label{eq:23-32} \\
 a_{32}a_{31}=a_{31}a_{32}  & a_{32}a_{31}=a_{31}a_{32} 
   \label{eq:32-31} \\
 a_{12}a_{41}=q a_{41}a_{12}+(1-q)a_{42}  & a_{12}a_{41}=a_{42}  & 
 a_{12}\prec a_{41} 
  \label{eq:12-41} \\
 a_{13}a_{41}=q a_{41}a_{13}+(1-q)Id  & a_{13}a_{41}=Id  & 
 a_{13}\prec a_{41}
  \label{eq:13-41} \\
  a_{22}a_{41}=q a_{41}a_{22}  & a_{22}a_{41}=0  &  a_{22}\prec a_{41} 
  \label{eq:22-41} \\
  a_{23}a_{41}=q a_{41}a_{23}  & a_{23}a_{41}=0  &  a_{23}\prec a_{41} 
  \label{eq:23-41} \\
  a_{31}a_{41}=q a_{41}a_{31}  & a_{31}a_{41}=0  &  a_{31}\prec a_{41} 
  \label{eq:31-41} \\
  a_{32}a_{41}=q a_{41}a_{32}  & a_{32}a_{41}=0  &  a_{32}\prec a_{41} 
  \label{eq:32-41} \\
  a_{12}a_{42}=a_{42}a_{12}  & a_{12}a_{42}=a_{42}a_{12}  
  \label{eq:12-42} \\
  a_{13}a_{42}=q a_{42}a_{13} +(1-q) a_{12}  & a_{13}a_{42}=a_{12}  & 
  a_{13}\prec a_{42} 
  \label{eq:13-42} \\
  a_{22}a_{42}=a_{42}a_{22}  & a_{22}a_{42}=a_{42}a_{22}    \label{eq:22-42} \\ 
 a_{23}a_{42}=q a_{42}a_{23}+(1-q) a_{22}  & a_{23}a_{42}=a_{22}  &
 a_{23}\prec a_{42} 
  \label{eq:23-42} \\
a_{42}a_{31}= a_{31}a_{42}+(1-q) a_{41}a_{32}   & a_{42}a_{31}=a_{31}a_{42}+a_{41}a_{32}    
  \label{eq:42-31} \\
 a_{32}a_{42}=q a_{42}a_{32}  & a_{32}a_{42}=0  & a_{32}\prec a_{42}
   \label{eq:32-42} \\
 a_{42}a_{41}=a_{41}a_{42}  & a_{42}a_{41}=a_{41}a_{42}
  \label{eq:42-41} \\
  \hline \\[-10pt]
\end{array}
\end{equation*}
}
\end{center}
\caption{
The hat algebra (\ref{eq:explicit-algebra}) for $(N,n)=(3,2)$.
The left and middle columns
correspond to the PASEP 
and TASEP cases, respectively.
In the right column
the ordering restricted by
the algebra of the TASEP case.
}\label{tab:32alg}
\end{table}
Let us consider the space of monomials
 generated by $a_{ij}$'s.
The TASEP algebra (the middle column of
table \ref{tab:32alg}) tells us the correct order
of the unknowns $a_{ij}$ in an arbitrary word 
consisting of these monomials.

For example, $a_{12}$ should be located to the right of  $a_{22}$
($a_{22}\succ a_{12}$) due to the relation $a_{12}a_{22}=0  $.
We listed the restriction for the reordering of 
the generators  in the right column of
table \ref{tab:32alg}.
One of the allowed orderings is 
\begin{equation}
   a_{41} \succ a_{31}\succ  a_{42}\succ
   a_{32}\succ  a_{22}\succ  a_{12}\succ
   a_{23}\succ  a_{13} .
\end{equation}
Let us consider the 
right-action of each generator $a_{ij}$
on the monomial (word)
\newcommand{\W}{\mathcal W}
\begin{equation}
   W  =   a_{41}^{n_{41}}  a_{31}^{n_{31}} 
   \cdots  a_{13}^{n_{13}}   \quad (  n_{ij}\ge 0   )   ,
\end{equation}
for the TASEP case.
We set
\begin{equation}
  a_{13} = \one^{\ot 7} \ot \del \, . 
\end{equation}
From the relation $a_{13}a_{23}=0 $,
one possibility for $a_{23}$ is
\begin{equation}
    a_{23} = \one^{\ot 6}
     \ot \del \ot A .
\end{equation}
From the first, forth and sixth relations,
 the action of $a_{12}$ is calculated as 
\begin{eqnarray}
\fl
   W  a_{12}
   = \left\{ \begin{array}{ll}
      a_{41}^{n_{41}} \cdots 
      a_{22}^{n_{22}} a_{12}a_{23}^{n_{23}}
      a_{13}^{n_{13}} +
      a_{41}^{n_{41}} \cdots   a_{32}^{n_{32}} 
      a_{22}^{n_{22}+1}a_{23}^{n_{23}-1}a_{13}^{n_{13}}
      & (n_{12}=0)  \\
      a_{41}^{n_{41}} \cdots   
      a_{22}^{n_{22}}a_{12}^{n_{12}+1}
        a_{23}^{n_{23}}a_{13}^{n_{13}} & (n_{12}\ge 1)   
   \end{array} \right.
\end{eqnarray}
Thus one can set
\begin{equation}
  a_{12} = \one^{\ot 5} \ot\del\ot\one^{\ot 2}
  +\one^{\ot 4}\ot\del\ot A\ot\eps\ot\del.
\end{equation}
In a similar way, we obtain conditions for the other monomials in the order 
$a_{22},\dots,a_{41}$, and we end up with
\begin{eqnarray}
  a_{22} = \one^{\ot 4}\ot\del\ot A\ot\one\ot A,\quad
  a_{32} =\one^{\ot 3}\ot\del\ot\one^{\ot 2}\ot A^{\ot 2},\\
  a_{42} = \one^{\ot 5}\ot\del\ot\one\ot\eps
               +\one^{\ot 4}\ot\del\ot A\ot\eps\ot\one
               +\one^{\ot 2}\ot\del\ot A\ot\one^{\ot 2}
                  \ot A^{\ot 2},\quad \\
  a_{31}=\one\ot\del\ot\one^{\ot 2}\ot A^{\ot 4}
   +\del\ot A\ot\eps\ot\del\ot A^{\ot 4}
   +\one^{\ot 3}\ot\del\ot\one\ot\eps\ot A^{\ot 2}, \\
  a_{41} = \one^{\ot 7}\ot\eps
   + \one^{\ot 2}\ot\del\ot A\ot\one\ot\eps\ot A^{\ot 2}
   +\del\ot A\ot\one\ot A^{\ot 5}.
\end{eqnarray} 

Accidentally (and fortunately),
the $q$-deformation of $\del$'s, $\eps$'s and $A$'s
in $a_{ij}$'s gives a solution to the hat relation
for general $q$.
Furthermore, the following simplification
still keeps the hat relation satisfied.
Erase the 1st, 2nd, 3rd, 4th and 5th
  components of the tensor products
  in each $a_{ij}$.
Erase the 3rd term in $a_{42}$,
 the 1st and 2nd terms in $a_{31}$
 and the 2nd and 3rd terms in   $a_{41}$.
Then we get a solution
\begin{equation}
\label{eq:32-rep}
\bfa^{(3,2)}=
\bordermatrix{
&{}_{1}  & {}_{2}     & {}_{3}    \cr
 \scriptstyle{1} & \one\ot\one\ot\one & A\ot\eps\ot\del+\del\ot\one\ot\one  & \one\ot\one\ot\del \cr
 \scriptstyle{2} & 0      & A\ot\one\ot A   & \one\ot\del\ot A \cr
 \scriptstyle{3} & \eps\ot A\ot A & \one\ot A\ot A  & 0   \cr
 \scriptstyle{4} &  \one\ot\one\ot\eps & A\ot\eps\ot\one+\del\ot\one\ot\eps & \one\ot\one\ot\one  }.
\end{equation}
It is  straightforward to verify that the 
 $a_{JK}$'s satisfy the algebra (\ref{eq:explicit-algebra}).
 Using this representation, one  can calculate  all matrix elements 
 (\ref{eq:def-psi-ele}) of the conjugation operator  $\psi_{\{1,2,3\},\{1,3\}}$ between 
the sectors  $\frs=\{1,2,3\}$ and  $\frt=\{1,3\}$. For example, one has 
\begin{eqnarray}
\eqalign{
\fl\quad
 \langle  1324 | \psi_{\{1,2,3\},\{1,3\}} | 2132 \rangle
 =\Tr \big( a^{(3,2)}_{12}  a^{(3,2)}_{31}  a^{(3,2)}_{23}  a^{(3,2)}_{42} \big)\\
\fl\quad
 =\Tr\big(
   (A\ot\eps\ot\del+\del\ot\one\ot\one) (\eps\ot A\ot A)
   (\one\ot\del\ot A) ( A\ot\eps\ot\one+\del\ot\one\ot\eps ) 
 \big) \\
\fl\quad
=\Tr(A\eps A) \Tr(\eps A\del\eps) \Tr(\del A^2)
 +\Tr(A\eps\del) \Tr(\eps A\del) \Tr(\del A^2\eps) \\
\fl\quad\quad
 +\Tr(\del\eps A) \Tr(A\del\eps) \Tr(A^2)
 +\Tr(\del\eps\del) \Tr(A\eps) \Tr(A^2\eps) \\
\fl\quad
=\frac{1+q^2}{(1-q^2)^2(1-q^3)}.
}
\end{eqnarray}
The full  elements of  matrix   $\psi_{\{1,2,3\},\{1,3\}}$
  are given by:
\begin{eqnarray}
\fl\quad\quad
\psi_{\{1,2,3\},\{1,3\}} =
{
\scriptsize
\left(
\begin{array}{cccccccccccc}
 a  & \bcsp \cdot  & \bcsp q b  & \bcsp \cdot  & \bcsp \cdot  & \bcsp q b  & \bcsp \cdot  & \bcsp c q  & \bcsp q d^3  & \bcsp \cdot  & \bcsp q d^3  & \bcsp q^2 e \\
 \cdot  & \bcsp a  & \bcsp b  & \bcsp \cdot  & \bcsp \cdot  & \bcsp \cdot  & \bcsp q b  & \bcsp q^2 d  & \bcsp q e  & \bcsp \cdot  & \bcsp q e  & \bcsp q d \\
 a  & \bcsp b  & \bcsp \cdot  & \bcsp b  & \bcsp c  & \bcsp \cdot  & \bcsp q^2 d  & \bcsp \cdot  & \bcsp \cdot  & \bcsp q^2 d  & \bcsp \cdot  & \bcsp q^2 e \\
 q b  & \bcsp a  & \bcsp \cdot  & \bcsp q e  & \bcsp b  & \bcsp q d  & \bcsp \cdot  & \bcsp \cdot  & \bcsp \cdot  & \bcsp q^2 d  & \bcsp q e  & \bcsp \cdot \\
 \cdot  & \bcsp q b  & \bcsp a  & \bcsp \cdot  & \bcsp q d  & \bcsp \cdot  & \bcsp c q  & \bcsp \cdot  & \bcsp q b  & \bcsp e  & \bcsp \cdot  & \bcsp q d \\
 b  & \bcsp \cdot  & \bcsp a  & \bcsp d  & \bcsp \cdot  & \bcsp c  & \bcsp \cdot  & \bcsp b  & \bcsp \cdot  & \bcsp e  & \bcsp d  & \bcsp \cdot \\
 \cdot  & \bcsp \cdot  & \bcsp \cdot  & \bcsp a  & \bcsp \cdot  & \bcsp b  & \bcsp \cdot  & \bcsp q^2 d  & \bcsp q e  & \bcsp q b  & \bcsp q e  & \bcsp q d \\
 \cdot  & \bcsp \cdot  & \bcsp \cdot  & \bcsp \cdot  & \bcsp a  & \bcsp \cdot  & \bcsp b  & \bcsp e  & \bcsp d  & \bcsp b  & \bcsp d  & \bcsp c \\
 \cdot  & \bcsp \cdot  & \bcsp \cdot  & \bcsp q b  & \bcsp q d  & \bcsp a  & \bcsp e  & \bcsp \cdot  & \bcsp \cdot  & \bcsp c q  & \bcsp q b  & \bcsp q d \\
 \cdot  & \bcsp \cdot  & \bcsp \cdot  & \bcsp q d^3  & \bcsp q b  & \bcsp q^2 e  & \bcsp a  & \bcsp \cdot  & \bcsp \cdot  & \bcsp c q  & \bcsp q d^3  & \bcsp q b \\
 \cdot  & \bcsp \cdot  & \bcsp \cdot  & \bcsp \cdot  & \bcsp q^2 e  & \bcsp \cdot  & \bcsp q^2 d  & \bcsp a  & \bcsp b  & \bcsp q^2 d  & \bcsp b  & \bcsp c \\
 \cdot  & \bcsp \cdot  & \bcsp \cdot  & \bcsp q e  & \bcsp \cdot  & \bcsp q d  & \bcsp \cdot  & \bcsp q b  & \bcsp a  & \bcsp q^2 d  & \bcsp q e  & \bcsp b \\
 q b  & \bcsp q e  & \bcsp q d  & \bcsp a  & \bcsp b  & \bcsp \cdot  & \bcsp q^2 d  & \bcsp \cdot  & \bcsp q e  & \bcsp \cdot  & \bcsp \cdot  & \bcsp \cdot \\
 c q  & \bcsp q b  & \bcsp q d  & \bcsp q b  & \bcsp a  & \bcsp q d  & \bcsp \cdot  & \bcsp e  & \bcsp \cdot  & \bcsp \cdot  & \bcsp \cdot  & \bcsp \cdot \\
 b  & \bcsp d  & \bcsp c  & \bcsp \cdot  & \bcsp \cdot  & \bcsp a  & \bcsp e  & \bcsp b  & \bcsp d  & \bcsp \cdot  & \bcsp \cdot  & \bcsp \cdot \\
 q^2 d  & \bcsp b  & \bcsp c  & \bcsp \cdot  & \bcsp \cdot  & \bcsp q^2 e  & \bcsp a  & \bcsp q^2 d  & \bcsp b  & \bcsp \cdot  & \bcsp \cdot  & \bcsp \cdot \\
 c q  & \bcsp q d^3  & \bcsp q b  & \bcsp q d^3  & \bcsp q^2 e  & \bcsp q b  & \bcsp \cdot  & \bcsp a  & \bcsp \cdot  & \bcsp \cdot  & \bcsp \cdot  & \bcsp \cdot \\
 q^2 d  & \bcsp q e  & \bcsp b  & \bcsp q e  & \bcsp q d  & \bcsp \cdot  & \bcsp q b  & \bcsp \cdot  & \bcsp a  & \bcsp \cdot  & \bcsp \cdot  & \bcsp \cdot \\
 \cdot  & \bcsp q d^3  & \bcsp q^2 e  & \bcsp \cdot  & \bcsp q b  & \bcsp \cdot  & \bcsp c q  & \bcsp \cdot  & \bcsp q d^3  & \bcsp a  & \bcsp \cdot  & \bcsp q b \\
 q^2 d  & \bcsp \cdot  & \bcsp q^2 e  & \bcsp b  & \bcsp \cdot  & \bcsp c  & \bcsp \cdot  & \bcsp q^2 d  & \bcsp \cdot  & \bcsp a  & \bcsp b  & \bcsp \cdot \\
 \cdot  & \bcsp q e  & \bcsp q d  & \bcsp \cdot  & \bcsp \cdot  & \bcsp \cdot  & \bcsp q^2 d  & \bcsp q b  & \bcsp q e  & \bcsp \cdot  & \bcsp a  & \bcsp b \\
 e  & \bcsp \cdot  & \bcsp q d  & \bcsp \cdot  & \bcsp \cdot  & \bcsp q d  & \bcsp \cdot  & \bcsp c q  & \bcsp q b  & \bcsp \cdot  & \bcsp q b  & \bcsp a \\
 q^2 d  & \bcsp q e  & \bcsp \cdot  & \bcsp q e  & \bcsp q d  & \bcsp b  & \bcsp \cdot  & \bcsp \cdot  & \bcsp \cdot  & \bcsp q b  & \bcsp a  & \bcsp \cdot \\
 e  & \bcsp d  & \bcsp \cdot  & \bcsp d  & \bcsp c  & \bcsp \cdot  & \bcsp b  & \bcsp \cdot  & \bcsp \cdot  & \bcsp b  & \bcsp \cdot  & \bcsp a
\end{array}\right)  },
\end{eqnarray}
where  we  have set 
\begin{eqnarray}
\fl\quad
a=\frac{1}{(1-q)^2 (1-q^2)},\  
b=\frac{1}{(1-q) (1-q^2)^2},\ 
c=\frac{1+q^2}{ (1-q^2)^2  (1-q^3)}, \\
\fl\quad
d=\frac{1}{(1-q^2)^2  (1-q^3)},\ 
e=\frac{1}{(1-q) (1-q^2)  (1-q^3)},
\end{eqnarray}
and  replaced 0 by  a dot  to make reading easier.
The ordering of the  bases is  lexicographic:
1234,1243,$\dots$,4321 for the sector  $\frs = \{1,2,3\}$,
and 1223,1232,$\dots$,3221 for $\frt = \{1,3\}$.
One can check explicitly that the conjugation relation
  $ \psi_{\{1,2,3\},\{1,3\}} M_{\{1,3\}} =M_{\{1,2,3\}}
   \psi_{\{1,2,3\},\{1,3\}}$ is satisfied.
We remark that all   the nonzero elements of this example have
a singularity at $q=1$ of order 3.

\subsection{The general  case}

One can construct a family of solutions to
(\ref{eq:hat-relation})
 for the general case $(N,n)$
 recursively by using the case $(N-1,n)$ if $n<N$
 or the case  $(N-1,n-1)$ if $n=N$:
\begin{equation}
\begin{array}{c}
  \nwarrow \quad\quad\,  \nwarrow \quad\quad\, 
  \nwarrow \quad\quad\, 
  \nwarrow \quad\, \nearrow \\
  (4,1) \quad (4,2) \quad (4,3) \quad (4,4) \\
  \nwarrow \quad\quad\, \nwarrow \quad\quad\, 
  \nwarrow \quad\, \nearrow \\
  (3,1) \quad (3,2) \quad (3,3)  \\
  \nwarrow \quad\quad\, \nwarrow \quad\, \nearrow \\
  (2,1)\quad (2,2)  \\
  \nwarrow \quad\, \nearrow \\
 (1,1)
\end{array}
\end{equation}
More explicitly,
for  $1\le n\le {N-1}$,
  $ a^{(N,n)}$ is defined in terms of  $ a^{(N-1,n)}$ by the following formula:

\newcommand{\bigbs}{\!\!\!\!\!\!\!\!\!\!\!\!\!\!\!\!\!\!\!\!\!}
\begin{eqnarray}
\label{eq:recursion}
\eqalign{\fl
   a^{(N,n)}_{JK} = \\
\fl
\left\{\begin{array}{ll}
\displaystyle     \sum_{J\le j\le n}
     a^{(N-1,n)}_{jK} \ot\one^{\ot(n-j)}\ot\eps\ot\one^{\ot(j-1-J)}\ot\del\ot A^{\ot(J-1)}
   &  (1\le J\le n,1\le K\le N-1),  \\[2mm]
    \one^{\ot(E-J)}\ot\del\ot A^{\ot(J-1)}
   & \bigbs (1\le J\le n ,K=N), \\[2mm]
     a^{(N-1,n)}_{JK}\ot A^{\ot n}
  & \bigbs (n+1\le J\le N,1\le K\le N-1), \\[2mm]
        0                 & \bigbs (n+1\le J\le N ,K=N),  \\[2mm]
\displaystyle  \sum_{1\le j\le n}
a^{(N-1,n)}_{jK}\ot\one^{\ot(n-j)}\ot\eps\ot\one^{\ot(j-1)}
        & \bigbs (J=N+1,1\le K\le N-1),  \\[2mm]
     \one^{\ot E}   & \bigbs (J=N+1,K=N), 
\end{array}\right. 
}
\end{eqnarray}
where $E=E(N,n)=nN-n^2+n-1$ is the number of 
the tensor products, and $\eps\ot\one^{\ot (-1)}\ot\del=\one$. 
For  $n=N$,
  $ a^{(N,N)}$ is defined in terms of  $ a^{(N-1,N-1)}$ as
\begin{eqnarray}
\label{eq:recursion-n=N}
   a^{(N,N)}_{JK} = 
\left\{\begin{array}{ll}
  \one^{\ot (N-1)} & (J=K=1),  \\  
  \del\ot a_{N,K-1}^{(N-1,N-1)} & (J=1,2\le K\le N), \\
  0 & (2\le J\le N,K=1),   \\
  A\ot a_{J-1,K-1}^{(N-1,N-1)} & (2\le J\le N,2\le K\le N), \\
  \eps\ot\one^{\ot(N-2)}   & (J=N+1,K=1),  \\
  \one\ot a_{N,K-1}^{(N-1,N-1)} & (J=N+1,2\le K\le N) .
\end{array}\right. 
\end{eqnarray}

We remark that in this representation the number of
tensor products grows typically  as $N^2$ (supposing that $n$ is of order $N$) whereas
 in the TASEP case
 the representation (\ref{eq:TASEP-solution}) 
  involves  only $(N-1)$-fold  tensor products.
Since the trace of each component 
of  the tensor product gives the factor $\frac{1}{1-q}$,
each nonzero element of $\psi$ 
has a singularity at $q=1$ of order $E(N,n)$.

This recursion has been obtained by 
investigating the case $(N,n)=(4,2)$
(see subsection \ref{subsec:32}) and 
 by  guessing the general $(N,n)$ case.
 The general  formulae (\ref{eq:recursion}) and (\ref{eq:recursion-n=N})
 are proved by verifying that 
 the  quadratic relations (\ref{eq:explicit-algebra})
 are satisfied.
This is  done by induction on the number of species  $N$.
I.e. one can check all the cases in the table  (\ref{eq:explicit-algebra})
assuming these are satisfied for $N\to N-1$.
The proof is not particularly illuminating,
which is in the same spirit as the one given in \cite{PEM}.

When $q=0$ and  $n=1,N$, the operators obtained from 
(\ref{eq:recursion}) and (\ref{eq:recursion-n=N})
  are identical to  the TASEP solution (\ref{eq:TASEP-solution}).
 This is not  true  anymore  for $ n\neq1,N$.
 In other words, in  the TASEP case ($q=0$), one has  two different  families of representations
  for the relations (\ref{eq:explicit-algebra}).
We expect, however, that the conjugation matrices
constructed by two  different  representations  are identical:
\begin{eqnarray}
   \psi_{\frs,\frs\setminus\{s_n\} } |_{q=0}
  = \widetilde\psi_{\frs,\frs\setminus\{s_n\} } 
\end{eqnarray}
for $2\le n\le N-1$ as well as for  $n=1,N$.
So far this identity has been checked for
small systems by using Mathematica.

\subsection{Conjugation paths through the Hasse diagram}

 We finally  consider the case of two  PASEP models with respectively  $N$  and $N'$ species
 and we suppose that   $N$  and $N'$  are  not consecutive integers. We show that
 a conjugation operator between the $N$-PASEP and the  $N'$-PASEP can be constructed by using recursively
 the results of the previous sections. 

 First, we define a $\star$-product on  operator valued matrices.
  Consider two operator valued matrices
$\A=\{A_{ij}\}_{ij}$ and $\mathcal B=\{B_{ij}\}_{ij}$ and 
let the symbol $\star$ denote the product 
\begin{eqnarray}
  \A \star \mathcal B = \{A_{ij}\ot B_{jk}\}_{ik} ,
\end{eqnarray}
which is {bilinear} and 
associative:
$(\A\star\mathcal B)\star\mathcal C
=\A\star(\mathcal B\star\mathcal C)$.
When $\A$ or $\mathcal B$ is a scalar-valued matrix,
$\star$ is just the usual product.
The following formula is satisfied:
\begin{equation}
  (\A\ot\mathcal B) \star (\mathcal C\ot\mathcal D)
   = (\A \star\mathcal C) \ot (\mathcal B\star\mathcal D) \, . 
\end{equation}

We now suppose that  we have found  solutions for the
 hat algebra for all $ N\in \N$:
\begin{eqnarray}\label{eq:MA-NN-1}
\fl\quad
 M^{(N)}_{\rm Loc}(\bfa^{(N)}\ot \bfa^{(N)}) - 
(\bfa^{(N)}\ot \bfa^{(N)})M^{(N-1)}_{\rm Loc}
= \bfa^{(N)} \ot\widehat{\bfa}^{(N)}-\widehat{\bfa}^{(N)}\ot \bfa^{(N)} \, .
\end{eqnarray}
Then, using the $\star$-product one can construct a solution of
 the  general hat-algebra defined as follows:
\begin{eqnarray}\label{eq:MA-NNprime}
 M^{(N)}_{\rm Loc}(\cX\ot \cX) = 
(\cX \ot \cX)M^{(N')}_{\rm Loc}
+ \cX \ot\widehat{\cX}-\widehat{\cX}\ot \cX  \, . 
\end{eqnarray}
 Indeed, one can show by induction that this relation is satisfied by the choice
\begin{eqnarray}\label{eq:A=aa}
   \cX = \bfa^{(N)} \star \bfa^{(N-1)} \star \cdots \star  \bfa^{(N'+1)} , \\
\label{eq:Ahat=sumaa}
   \widehat \cX = \sum_{N\ge i \ge N'+1}
    \bfa^{(N)} \star  \cdots  \star \bfa^{(i+1)} \star  \widehat \bfa^{(i)}
    \star    \bfa^{(i-1)} \star  \cdots  \star \bfa^{(N'+1)}  \, . 
\end{eqnarray}

Generically there is no  linear relation
of the type~(\ref{eq:ahat=diaga})  
 relating $\cX$ and  $\widehat \cX$ ; 
hence, a closed algebra 
 cannot be defined by  the elements of $\cX$  alone  (except for the $N'=N-1$ case).
Furthermore, we observe that for $1\le n<n'\le N$,
$a^{(N,n')}\star a^{(N-1,n)} \neq 
a^{(N,n)}\star a^{(N-1,n'-1)}$, 
and  therefore $\cX$
depends on the choice of the intermediate values of the $n$'s.
In other words,
one has several solutions to (\ref{eq:MA-NNprime}).
We conjecture, however,
for a $N$-species sector
$\frs=\{s_1,\dots,s_N\}\ (s_1<\cdots<s_N,
1\le n<n'\le N)$,
the ``commutation relation''
holds up to an overall factor as
\begin{equation}
\label{eq:commu}
\fl\quad
   \psi_{  \frs,\frs\setminus\{s_n\}  } 
   \psi_{  \frs\setminus\{s_n\},\frs\setminus\{s_n,s_{n'}\}  } 
= \frac{\prod_{i=n'+1}^{N} (1-q^{s_i-s_n})   }
         {\prod_{i=1}^{n-1} (1-q^{s_{n'}-s_i})  }
 \psi_{  \frs,\frs\setminus\{s_{n'}\}  } 
 \psi_{  \frs\setminus\{s_{n'}\},\frs\setminus\{s_n,s_{n'}\}  } ,
\end{equation}
\newcommand{\longnearrow}{\mbox{\rotatebox[origin=c]{225}{
$\longleftarrow\!\!\! -$}}}
\newcommand{\longnwarrow}{\mbox{\rotatebox[origin=c]{315}{
$\longleftarrow\!\!\! -$}}}
or equivalently the following diagram commutes
 up to an overall factor:
\begin{equation}
\begin{array}{rcl}
  &    V_\frs   &    \\
  \psi_{\frs,\frs\setminus\{s_n\}}\!\!\!
    \longnearrow  & &
    \longnwarrow \!\!\!
  \psi_{\frs,\frs\setminus\{s_{n'}\}}   \\ 
 V_{\frs\setminus\{s_n\}}\quad \quad\quad & & 
  \quad\quad\  V_{\frs\setminus\{s_{n'}\}}  \\
  \psi_{\frs\setminus\{s_n\},\frs\setminus\{s_n,s_{n'}\}}\!\!\!    \longnwarrow  & &
    \longnearrow \!\!\!
  \psi_{\frs\setminus\{s_{n'}\},\frs\setminus\{s_n,s_{n'}\}}   \\  
      &  \!\!\!\!\!\!\!\!\!
       V_{\frs\setminus\{s_n,s_{n'}\}}\!\!\!\!\!\!\!\!\!  & 
\end{array}.
\end{equation}
So far this identity has been checked for
small systems by using Mathematica.

For $N'=1$ the relation (\ref{eq:MA-NNprime}) corresponds to
the usual matrix Ansatz for the stationary state
(which is unique up to an overall constant in each sector),
and one has several matrix representations for the
stationary state according to the path
$\emptyset\to\cdots\to\frs$
on the Hasse diagram.
The stationary state
was first obtained in \cite{PEM} , where 
the path corresponds to
$\emptyset\to\{s_1\}\to\{s_1,s_2\}\to\cdots\to\frs\setminus\{s_N\}\to\frs$.

\hfill\break
{\it Remark:}  In the symmetric exclusion case   $p=q=1$ (SSEP),
  the Markov matrix obeys the detailed-balance condition,
  and is a symmetric matrix: $M^{\rm T} = M$.
Thus the system in each sector converges
  to an equilibrium stationary state,
  where all the possible configurations
  are realized with an equal probability.
 Since the conjugation matrix 
  $\psi$ has singularity $(1-q)^{-E(N,n)}$,
  we take the limit $q\to 1$
  after multiplying it by $(1-q)^{E(N,n)}$:
\begin{equation}
   \overline\psi_{\frs,\frs\setminus\{s_n\}}=
   \lim_{q\to 1}(1-q)^{E(N,n)}\psi_{\frs,\frs\setminus\{s_n\}}
\end{equation}
for $\frs=\{s_1,\dots,s_N\}$ and $1\le n\le N$.
This matrix satisfies
 the conjugation relation  (\ref{eq:main-relation}):
$  M_{\frs}\overline\psi_{\frs,\frs\setminus\{s_n\}} =
  \overline\psi_{\frs,\frs\setminus\{s_n\}} M_{\frs,\setminus\{s_n\}} .$
 In the SSEP case,
  one can easily find a few other possibilities for 
  a conjugation matrix satisfying (\ref{eq:main-relation}).
 For example, the transpose of the identification operator $\varphi$
also satisfies 
$   M_{\frs}\varphi^{\rm T}_{\frt\frs} =
  \varphi^{\rm T}_{\frt\frs} M_{\frt}$  
for $\frs\supset\frt$
since $  M_{\frs}^{\rm T} =  M_{\frs}$
and $  M_{\frt}^{\rm T} =  M_{\frt}$.
 Another  simple solution 
  (one dimensional representation)
  to the hat relation (\ref{eq:hat-relation})
  with (\ref{eq:ahat=diaga}) is  
\begin{equation}
   a'_{JK} =
     \left\{\begin{array}{ll}
       0 &  
       \begin{array}{l}
       ( 2\le J\le n, 1 \le K\le J-1  \\
               \ {\rm or}\ n+1\le J\le N, J \le K\le N  ),
          \end{array}     \\
       1 &  (\rm otherwise).
   \end{array}\right.
\end{equation}
Then  the matrix $\psi'$ defined as
$ \langle J_1\cdots J_L | \psi'_{\frs\frt}
 | K_1\cdots K_L \rangle
 = \prod_{1\le i\le L}  a'_{J_iK_i}$ 
also satisfies 
$  M_{\frs}\psi'_{\frs\frt} = \psi'_{\frs\frt} M_{\frt} .$

\section{The $N$-ASEP as a  multistate vertex model}
\label{sec:Perk}

\begin{figure}
   \includegraphics[height=9cm]{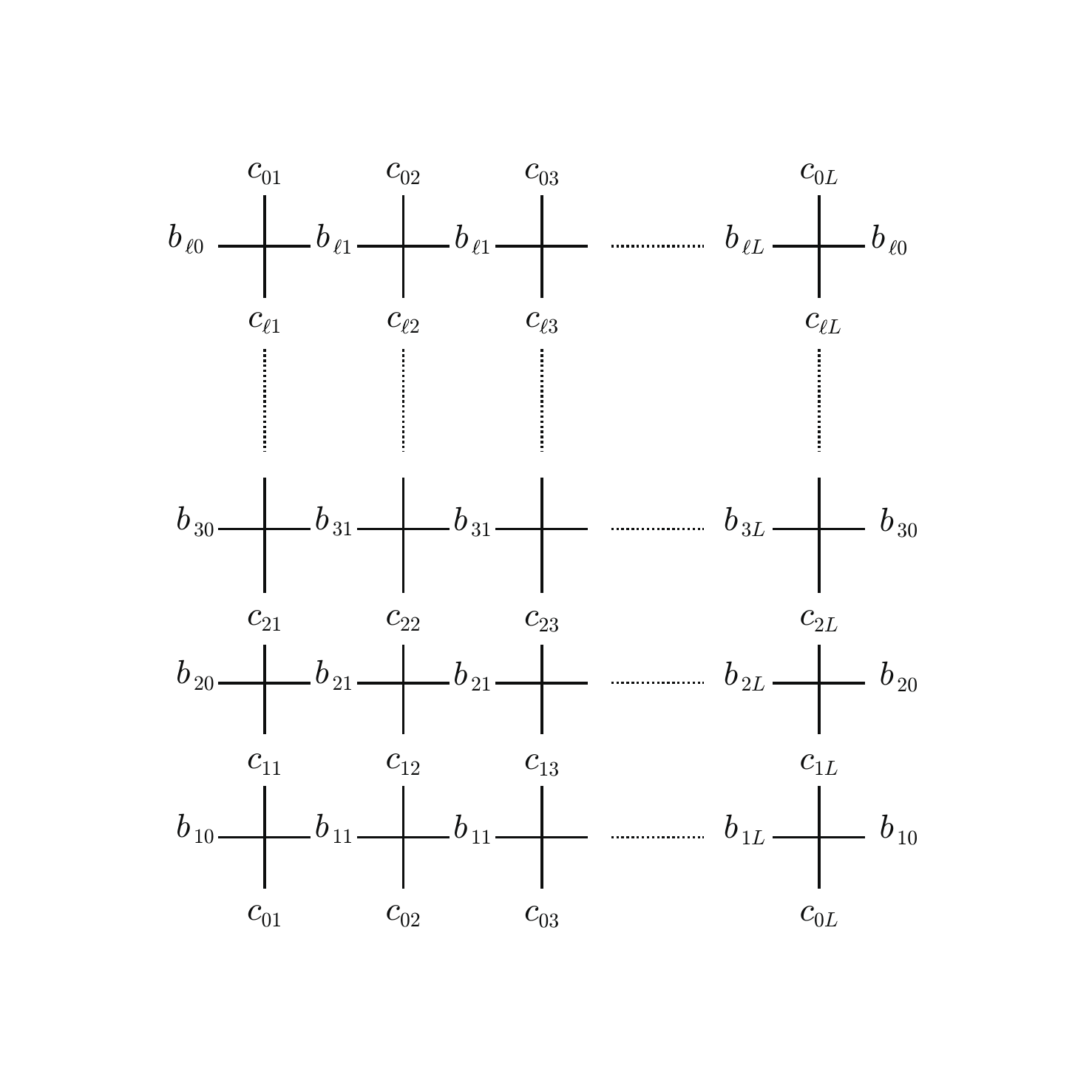}
      \caption{The two-dimensional vertex model.
The $N$-ASEP corresponds to the case where Boltzmann weights are given as equation (\ref{eq:weight}).}
 \label{fig:vertex-model}
 \end{figure}

Let us consider the two dimensional vertex model,
e.g. on $\ell\times L$ lattice.
As in figure \ref{fig:vertex-model}, 
each vertex $(i,j)$
  has a Boltzmann weight $W_{i,j}$
 defined by values 
  $(b_{i,j-1},\, c_{i-1,j},\, b_{i,j},\, c_{i,j})$ that can be assigned to its four connected edges.
The partition function is 
given by
\begin{equation}
Z=\displaystyle
\sum_{\rm configuration}\ 
\prod_{{\rm vertices} {\;(i,j)}} W_{i,j} \, .
\end{equation}

The Perk-Schultz model defines
a family of vertex models that
 are exactly solvable \cite{PS}. 
 As we will  shortly review, the $N$-ASEP can be realized as a special
 case of the  Perk-Schultz models for a particular choice of the vertex
 weights. We  will also 
 investigate the conjugation relation for 
{this particular restriction of} the Perk-Schultz model.

Define a matrix
$R(\lambda)\in{\rm End}\left(\left(\C^{N+1}\right)^{\ot2}\right)$ as
\begin{eqnarray}
R(\lambda)=\rho\left(1+\lambda M^{(N)}_{\rm Loc} \right),
\end{eqnarray}
where $\rho$ is the permutation matrix:
$\rho(|\alpha\rangle\ot|\beta\rangle)
=|\beta \rangle\ot|\alpha \rangle$.
By regarding each element
$R_{xy}^{zw}(\lambda)=
\langle x y|  R(\lambda) | zw \rangle$
as the Boltzmann weight of each vertex, the  $N$-ASEP
corresponds to  a special case of 
 the Perk-Schultz model with 
 $N+1$ states,
 where the non-zero elements  are given as 
\begin{equation}
\label{eq:weight}
\begin{array}{ccc}
 \alpha &\alpha &\beta  \\   
 \alpha\cross\alpha &  
 \alpha\cross\beta &  
 \alpha\cross\alpha \\  
 \alpha & \beta &\beta  \\
 R^{\alpha\alpha}_{\alpha\alpha}(\lambda)=1,  &
 R^{\alpha\beta}_{\alpha\beta}(\lambda)=\lambda\Theta(\alpha-\beta),
 &
 R^{\alpha\beta}_{\beta\alpha}(\lambda)=1-\lambda\Theta(\alpha-\beta),
 \end{array}
\end{equation} 
see \cite{AKSS,PS,Sc}.
The matrix $R(\lambda)$ satisfies the Yang-Baxter relation
\begin{eqnarray}\label{eq:RRR}
  R_{bc}(\nu) R_{ac}(\mu) R_{ab}(\lambda)
  =R_{ab}(\lambda) R_{ac}(\mu) R_{bc}(\nu)
\end{eqnarray}
with $\lambda=\frac{\mu-\nu}{1-(p+q)\nu+p q \mu \nu}$.
The indices $a,b$ and $c$ specify the spaces on which the $R$ matrices act.
The monodromy matrix is defined as
\begin{eqnarray}
\T_a(\lambda)=R_{aL}(\lambda)\cdots R_{a1}(\lambda)
\end{eqnarray}
acting on $\C^{N+1}\ot
\left(\C^{N+1}\right)^{\ot L}$ 
where the first space is the so-called auxiliary space
and is  denoted by $a$. It satisfies
the ``global Yang-Baxter relation''
\begin{eqnarray}\label{eq:gYBE}
\T_b(\nu)\T_a(\mu)R_{ab}(\lambda)
=R_{ab}(\lambda)\T_a(\mu)\T_b(\nu)
\end{eqnarray}
with $\lambda=\frac{\mu-\nu}{1-(p+q)\nu+p q \mu \nu}$.
The row-to-row
transfer matrix $T(\lambda)$ is defined by
the trace of the monodromy matrix
over the auxiliary space,
\begin{equation}
T(\lambda)={\rm Tr}_{a} \T_a(\lambda).
\end{equation}
{The Yang-Baxter relation implies that it} constitutes a one-parameter commuting family:
\begin{equation}
[T(\lambda_1) ,T(\lambda_2)]=0.
\end{equation}
The Markov matrix can be rewritten in terms of
$T(\lambda)$ as
\begin{eqnarray}
M^{(N)}=\left.
\frac{\rm d}{{\rm d}\lambda}\log T(\lambda)
\right|_{\lambda=0} ,
\end{eqnarray}
which indeed satisfies $[M^{(N)},T(\lambda)]=0$.
The transfer matrix $T(\lambda)$
and the Markov matrix  $M^{(N)}$
can be diagonalized 
by using the Bethe Ansatz \cite{AB,AKSS,Baxter,Sc}.

The number of each species of particles is invariant under
the action of the transfer matrix $T(\lambda)$
as well as the Markov matrix
 ($T(\lambda)V_\frs \subseteq V_\frs$)
 and we denote the restriction  on the sector $\frs$ by $T_\frs(\lambda)$.
We conjecture that the transfer matrix also satisfies the conjugation relations with the identification matrix 
and  with the new conjugation matrix
up to order $\lambda^L$:
for sectors $\frs = \{s_1<\cdots<s_N\}$ and $\frt = \frs\setminus\{s_n\}$
\begin{eqnarray}
\vp_{\frt\frs} T_\frs(\lambda) 
-T_\frt(\lambda) \vp_{\frt\frs}
  = q^{s_n}\lambda^L
  \vp_{ \frt\frs} ,\\
 T_\frs(\lambda) \psi_{\frs\frt}
-\psi_{\frs\frt}T_\frt(\lambda)  
  = q^{s_n}\lambda^L\psi_{\frs\frt}.\end{eqnarray}

\hfill\break
{\it Remark:}
In the nested algebraic Bethe Ansatz technique, 
 eigenvectors of $M^{(N)}$
 are constructed
 by the action of a product of
  ``$B$-operators'' (elements
 of $\mathcal  T(\lambda)$) 
on the vector  $|1\cdots 1\rangle $. 
On the other hand,
the generalized matrix Ansatz also 
enables us to construct an  eigenvector
 by the action of a product of $\psi$'s 
  on  an {\it eigenvector}
with the same eigenvalue
{\it  in a lower sector}.
In particular,  the stationary state 
can be written as
 $\psi\cdots \psi |1\cdots 1\rangle $.
It seems that our conjugation operator 
 and the $B$-operator play similar roles.
However the generalized matrix Ansatz never gives us information
about eigenvectors with new eigenvalues in each sector.

\section{Conclusion}
\label{sec:conclusion}

We have applied the generalized matrix Ansatz to the multispecies ASEP, where the central issue is reduced to finding a representation 
 for the hat algebra.
The family of representations that we find
 here is defined by a recursion formula with respect to 
the number of species $N$.
This new solution is not obtained by perturbation of the known
 solution for the TASEP.
 The SSEP is also a special case where
a one-dimensional representation exists.  
We conjecture
that the conjugation relation  continues to  hold  (modulo $\lambda^L$)
for the Perk-Schultz transfer matrix.

As we remarked in the last section,
our generalized matrix Ansatz 
does not enable us to construct 
eigenvectors with new eigenvalues in each sector, which is
 a problem to be  solved.
Another open question is 
whether there exists another family of representations to the hat algebra.
(In fact we have found different 
families of representations for the TASEP and SSEP cases.)
Applying  the technique to open boundary conditions   
 with injection and extraction of particles
 or other driven-diffusive processes
 with rules more general  
  than equation (\ref{eq:rate}) 
 can also be  an interesting  study
 \cite{BE,Cantini}.

\quad

\ack{ It is a pleasure to  thank C. Banderier, M. Bauer and G.
  Duchamp  for very useful discussions
 on the deformations of quadratic algebras.
We also thank S. Mallick for a critical reading of the manuscript.
C. Arita is a JSPS Fellow for Research Abroad.}

\quad

\end{document}